\documentclass[prd,superscriptaddress,preprint]{revtex4}

\usepackage{mathrsfs,amsmath}
\usepackage{eurosym}
\usepackage{amssymb}
\usepackage{bbold}

\usepackage{latexsym,epsfig,epstopdf}
\usepackage{amssymb,amsmath,amsthm}

\usepackage{times}
\usepackage{color}

\usepackage{lmodern}
\usepackage[utf8]{inputenc}

\usepackage{graphicx}

\usepackage{color}

\usepackage{braket}

\usepackage[mathscr]{euscript} 

\def\be{\begin{equation}}
\def\ee{\end{equation}}
\def\bea{\begin{eqnarray}}
\def\eea{\end{eqnarray}}

\begin{document}

\title{Quantum Hall admittance in non-Hermitian systems}

\author{Annan Fan}
\affiliation{School of Physics, Sun Yat-Sen University, Guangzhou 510275, China}

\author{Guang-Yao Huang}
\affiliation{Institute for Quantum Information and State Key Laboratory of High Performance Computing,
College of Computer, National University of Defense Technology, Changsha 410073, China}

\author{Shi-Dong Liang}
\email{stslsd@mail.sysu.edu.cn}
\affiliation{School of Physics, Sun Yat-Sen University, Guangzhou 510275, China}
\affiliation{State Key Laboratory of Optoelectronic Material and Technology, and\\
Guangdong Province Key Laboratory of Display Material and Technology,
Sun Yat-Sen University, Guangzhou, 510275, China}

\begin{abstract}
We propose a pair of the complex Berry curvatures associated with the non-Hermitian Hamiltonian and its Hermitian adjoint
to reveal new physics in non-Hermitian systems. We give the complex Berry curvature and Berry phase for the two-dimensional non-Hermitian Dirac model. The imaginary part of the complex Berry phase induces the quantum Hall susceptance such that the quantum Hall conductance is generalized to quantum Hall admittance for non-Hermitian systems, which implies that the non-Hermiticity of systems could induce an intrinsic quantum capacitance or inductance of systems depending on the non-Hermitian parameters. We analyze the complex energy band structures of the two-dimensional non-Hermitian Dirac model and demonstrate the point and line gaps
and their closings as the exceptional degeneracy of the energy bands in the parameter space, which are associated with topological phases. In the continuum limit, we obtain the complex Berry phase in the parameter space.


\end{abstract}

\maketitle
\tableofcontents

\section{Introduction}

Geometric and topological descriptions of quantum systems provide powerful mathematical tools to reveal new quantum phases in condensed matter physics.\cite{Kane,Hasan,Tkachov, Qi,Wen,Zohar,Andrei} The quantum Hall effect and quantum spin Hall effects can be understood in terms of the topological Chern classes, $\mathbb{Z}$ and $\mathbb{Z}_{2}$.\cite{Tkachov,Schnyder,Ryu,Chiu} These discoveries have inspired many attempts to study the topological phase or states of condensed matter systems, \cite{Chiu,Ryu} which can be classified by ten topological classes, based on the symmetry group and homological invariance.\cite{Chiu} The design of robust topological states promises potential applications in quantum materials and technology.\cite{Ramy,Zhang,Wehling,Tang}

Recently, non-Hermitian systems have attracted growing attention, which exhibit unconventional characteristics and potential applications.\cite{Ramy,Tang,Fidkowski,Hu,Leykam,Martinez}
Many efforts have been devoted to exploring topological states in non-Hermitian systems.\cite{Liang,Fu,Jiang,Gong,Kohei}
The non-Hermiticity of systems appear in non-equilibrium open systems with dissipative phenomena, such energy gain or loss and non-conserved probability of electrons.
The non-Hermitian matrix representation of quantum mechanics involves some unconventional characteristics beyond standard quantum mechanics based on Hermitian operators, such as complex energy spectra and their nonorthogonal eigenvectors.
In general, problems addressed include
(1) How to construct a non-Hermitian generalization of the canonical (Hermitian) quantum formalism by introducing the $\eta-$inner product, pseudo-Hermiticity and non-unitary similarity transformation to reformulate the framework of quantum mechanics. \cite{Bender,Ali1,Moiseyev,Ghosh}
(2) How to generalize the classification of the topological phases of Hermitian systems to non-Hermitian systems based on
the Altland-Zirnbauer (AZ) symmetry and their generalization for non-Hermitian systems.\cite{Gong,Kohei}

Very recently, one proposed a complete topological classification for non-Hermitian systems based on symmetry and topology,
which generalize ten-fold topological classes of Hermitian systems to 38-fold symmetry for non-Hermitian systems, including generalized AZ symmetry, sublattice symmetry and pseudo-Hermiticity, which covers all the internal symmetries for non-Hermitian systems. \cite{Gong,Kohei}
The topological classes are characterized by winding number, vorticity, Berry curvature,and Chern number of their quantum states. \cite{Gong,Kohei,Fu,Jiang,Prodan}
The vorticity defined in the complex-energy plane is demonstrated to be equivalent to winding number for classification of topological invariant. \cite{Kohei,Fu}
In particular, one found that the complex energy band gap for non-Hermitian systems could form a point or line gap to preserves its topological invariant under a unitary or Hermitian flattening transformation. \cite{Kohei} The energy band gaps closing to form exceptional points and lines as a reference point or line associate with topological phases.\cite{Kohei,Gong} The non-Hermiticity of systems deforms the Bloch-wave behavior to yield the skin effect for lattice models.\cite{Lee,Alvarez,Kawabata,Liu}
One introduced the biorthogonal polarization to modify the conventional bulk-boundary correspondence to show the zero modes,\cite{Kunst,Yao,Esaki} and the topological edge states,  finite-size effects in the non-Hermitian Su-Schrieffer-Heeger (SSH) model.\cite{Zhu,Jiang,Dangel,Lieu,Chen}
In particular, Chen and Zhai studied the Hall conductance of a non-Hermitian Chern insulator and found some deviations of the quantized Chern number due to the non-Hermitian effects.\cite{Yu}

In general, non-Hermiticity of systems implies that the energy bands are in general complex such that the gaps and their corresponding exceptional degeneracy could induce some new and more rich phenomena beyond the Hermitian systems. A natural question is what role play the Hermitian adjoint of the Hamiltonian in non-Hermitian systems.
What physical observables and phenomena come from non-Hermiticity of the non-Hermitian systems beyond the Hermitian systems?

In this paper, we will give the complex Berry curvature and Berry phase for the two-dimensional (2D) non-Hermitian Dirac model. In Sec. II we will redefine
a pair of complex Berry curvatures from the non-Hermitian Hamiltonian and its Hermitian adjoint to reveal some new features of quantum states. We will generalize quantum Hall conductance to quantum Hall admittance, which is given by the complex Berry for non-Hermtian systems in Sec. III.
The quantum Hall admittance implies that the non-Hermiticity of systems induces an intrinsic Hall susceptance, which includes the electric capacitive and inductive properties depending on the non-Hermitian parameters. In Sec. IV, we will analyze the energy band structure of the two-dimensional (2D) non-Hermitian Dirac model, giving the point and line gap exceptional degeneracies and their associated Dirac cone and Weyl node, which imply the existence of the topological phase. In the continuum limit we will give the complex Berry phase of the 2D non-Hermitian Dirac model in the parameter space. The conclusion and discussion will be given in Sec. V. In the appendix we will give the derivation of the complex Berry curvature and Berry phase for the 2D non-Hermitian systems.

\section{Complex Berry curvature and Berry phase in Non-Hermitian two-band model} \label{SEC_SIMPLEST}
Let us consider a non-Hermitian two-band model with translation invariance. The effective Hamiltonian is described by the 2D non-Hermitian Dirac model in momentum space,\cite{Fu}
\begin{equation}\label{H1}
\mathcal{H}(\mathbf{k})=\mathbf{h}(\mathbf{k})\cdot \mathbf{\sigma},
\end{equation}
where $\mathbf{h}(\mathbf{k})$ is a complex function, where $\mathbf{k}$ is the generalized crystal momentum in the complex domain.
$\mathbf{h}(\mathbf{k})$ may be regarded as a generalized Zeeman-like magnetic field. $\mathbf{\sigma}$ is the spin 1/2 Pauli operator.
Thus, the Hamiltonian is not Hermitian, $\mathcal{H}^{\dagger}(\mathbf{k})=\mathbf{h}^{*}(\mathbf{k})\cdot \mathbf{\sigma}\neq\mathcal{H}(\mathbf{k})$, where $*$ is the complex conjugated operator.
The eigen equations of the Hamiltonian and its Hermitian adjoint are given

\begin{eqnarray}\label{EEq1}
\mathcal{H}|\psi_{\pm}^{R}\rangle &=& E_{\pm}|\psi_{\pm}^{R}\rangle \\
\mathcal{H}^{\dagger}|\phi_{\pm}^{L}\rangle &=& E_{\pm}^{*}|\phi_{\pm}^{L}\rangle,
\end{eqnarray}

where the eigen vectors of $\phi_{\pm}^{L}$ and $\psi_{\pm}^{R}$ form a biorthonormal basis, $\{\phi_{\pm}^{L}, \psi_{\pm}^{R}\}$.
The completeness relations for $\mathcal{H}$ and $\mathcal{H}^{\dagger}$ can be written as

\begin{eqnarray}\label{CR2}
I &=& P_{+}+P_{-}\equiv|\psi_{+}^{R}\left\rangle\right\langle \phi_{+}^{L}|+|\psi_{-}^{R}\left\rangle\right\langle \phi_{-}^{L}| \\
I &=& P_{+}^{\dagger}+P_{-}^{\dagger}\equiv|\phi_{+}^{L}\left\rangle\right\langle\psi_{+}^{R} |+|\phi_{-}^{L}\left\rangle\right\langle \psi_{-}^{R}|,
\end{eqnarray}

where $P_{\pm}$ and $P_{\pm}^{\dagger}$ are projection operators for $\mathcal{H}$ and $\mathcal{H}^{\dagger}$.

The spectral representations of $\mathcal{H}$ and $\mathcal{H}^{\dagger}$ are given by

\begin{eqnarray}\label{HH}
\mathcal{H}=E_{+}|\psi_{+}^{R}\left\rangle\right\langle \phi_{+}^{L}|+E_{-}|\psi_{-}^{R}\left\rangle\right\langle \phi_{-}^{L}| \\
\mathcal{H}^{\dagger}=E_{+}^{*}|\phi_{+}^{L}\left\rangle\right\langle\psi_{+}^{R} |+E_{-}^{*}|\phi_{-}^{L}\left\rangle\right\langle \psi_{-}^{R}|,
\end{eqnarray}
Using the spectral representation of the Hamiltonian (\ref{HH}), the projection operators can be rewritten as

\begin{eqnarray}\label{PP2}
P_{\pm}&=&\frac{1}{2}\left(I\pm \widehat{\mathbf{h}}\cdot \mathbf{\sigma}\right) \\
P_{\pm}^{\dagger}&=&\frac{1}{2}\left(I\pm \widehat{\mathbf{h}^{*}}\cdot \mathbf{\sigma}\right),
\end{eqnarray}
where $\widehat{\mathbf{h}}:=\frac{\mathbf{h}}{h}$ is the dimensionless crystal momentum.

Let us consider an adiabatic evolution of the system, the generalized Berry connections are defined by\cite{Liang,Keck} (see Appendix A)

\begin{eqnarray}\label{AA2}
A_{\pm} &=& i\left\langle \phi_{\pm}^{L}|d|\psi_{\pm}^{R}\right\rangle \\
A_{\pm}^{\dagger} &=& i\left\langle \psi_{\pm}^{R}|d|\phi_{\pm}^{L}\right\rangle,
\end{eqnarray}
where $d$ is the exterior differential operator.
For the 2D parameter space denoted by $(k_x,k_y)$,
the generalized Berry curvatures for non-Hermitian systems are defined as $\Omega_{\pm}=d A_{\pm}=\Omega_{\pm}^{z}dk_{x}\wedge dk_{y}$ and
$\Omega_{\pm}^{\dagger}=d A_{\pm}^{\dagger}=\Omega_{\pm}^{z\dagger}dk_{x}\wedge dk_{y}$ respectively, where

\begin{eqnarray}\label{BC2}
\Omega_{\pm}^{z} &=& i\left[\langle\partial_{k_x}\phi_{\pm}^{L} |\partial_{k_y}\psi_{\pm}^{R}\rangle-\langle \partial_{k_y}\phi_{\pm}^{L} |\partial_{k_x}\psi_{\pm}^{R}\rangle\right] \\
\Omega_{\pm}^{z\dagger} &=& i\left[\langle\partial_{k_x}\psi_{\pm}^{R} |\partial_{k_y}\phi_{\pm}^{L}\rangle-\langle \partial_{k_y}\psi_{\pm}^{R} |\partial_{k_x}\phi_{\mp}^{L}\rangle\right],
\end{eqnarray}

Note that $\partial_{k_\alpha}P_{\pm}=|\partial_{k_\alpha}\psi_{\pm}^{R}\left\rangle\right\langle \phi_{\pm}^{L}|+ |\psi_{\pm}^{R}\left\rangle\right\langle \partial_{k_\alpha}\phi_{\pm}^{L}|$,
the Berry curvature in (\ref{BC2}) can be expressed as(see Appendix A)

\begin{eqnarray}\label{BC3}
\Omega_{\pm}^{z} &=& i\textrm{Tr}\left(P_{\pm}\left[\partial_{k_x}P_{\pm},\partial_{k_y}P_{\pm}\right]\right) \\
\Omega_{\pm}^{z\dagger} &=& i\textrm{Tr}^{\dagger}\left(P_{\pm}^{\dagger}\left[\partial_{k_x}P_{\pm}^{\dagger},\partial_{k_y}P_{\pm}^{\dagger}\right]\right),
\end{eqnarray}
where the traces are defined as $\textrm{Tr}(\bullet):=\sum_{n}\langle\phi_{n}^{L}|\bullet|\psi_{n}^{R}\rangle$ and  $\textrm{Tr}^{\dagger}(\bullet):=\sum_{n}\langle\psi_{n}^{R}|\bullet|\phi_{n}^{L}\rangle$ for the biorthnormal basis.

The generalized Berry curvature for the non-Hermitian cases can then be expressed as (see Appendix A)

\begin{eqnarray}\label{BC4}
\Omega_{\pm}^{z} &=& \mp\frac{1}{2}\widehat{\mathbf{h}}\cdot
\left(\frac{\partial \widehat{\mathbf{h}}}{\partial k_{x}}\times
\frac{\partial \widehat{\mathbf{h}}}{\partial k_{y}}\right) \\
\Omega_{\pm}^{z\dagger}&=&\mp\frac{1}{2}\widehat{\mathbf{h}}^{*}\cdot
\left(\frac{\partial \widehat{\mathbf{h}}^{*}}{\partial k_{x}}\times
\frac{\partial \widehat{\mathbf{h}}^{*}}{\partial k_{y}}\right).
\end{eqnarray}

They have the same form as the Hamiltonian of the Hermitian 2D Dirac model, except that the real $\mathbf{h}$ is replaced by the complex $\mathbf{h}$.
Consequently, we redefine the Berry curvature for non-Hermitian systems by $\Omega_{\pm}^{z}=\frac{1}{2}\left(\Omega_{\pm}^{z,r}+i\Omega_{\pm}^{z,i}\right)$, where the real and imaginary parts are given by

\begin{eqnarray}\label{BC5}
\Omega_{\pm}^{z,r} &=& \Omega_{\pm}^{z}+\Omega_{\pm}^{z\dagger}=\Omega_{\pm}^{z,r} \\
\Omega_{\pm}^{z,i} &=& \Omega_{\pm}^{z}-\Omega_{\pm}^{z\dagger}=\Omega_{\pm}^{z,i},
\end{eqnarray}
which include all effects of the non-Hermitian Hamiltonian and its Hermitian adjoint. The Berry curvature for non-Hermitian systems is associated with different physical phenomena. Both real and imaginary parts are quantized to associate with the topological phase.  Thus, the Berry phase for the non-Hermitian Dirac model can be also generalized to the complex domain, $\gamma_{\pm}=\gamma_{\pm}^{R}+i\gamma_{\pm}^{I}$, where

\begin{eqnarray}\label{CN1}
\gamma_{\pm}^{R} &=& \mp\frac{1}{4\pi}\Re \int_{\textrm{BZ}} \widehat{\mathbf{h}}\cdot
\left(\frac{\partial \widehat{\mathbf{h}}}{\partial k_{x}}\times
\frac{\partial \widehat{\mathbf{h}}}{\partial k_{y}}\right)dk_{x}dk_{y} \\
\gamma_{\pm}^{I} &=& \mp\frac{1}{4\pi}\Im \int_{\textrm{BZ}} \widehat{\mathbf{h}}\cdot
\left(\frac{\partial \widehat{\mathbf{h}}}{\partial k_{x}}\times
\frac{\partial \widehat{\mathbf{h}}}{\partial k_{y}}\right)dk_{x}dk_{y}.
\end{eqnarray}

It can be seen that both of the real and imaginary Berry phase are quantized and associated with some non equilibrium phenomena in non-Hermitian systems, which is equivalent  to the complex winding number in the 1D non-Hermitian two-band model.\cite{Jiang}
What is new physics of the complex Berry curvature and complex Berry phase for non-Hermitian systems?

\section{Quantum Hall admittance}
In order to explore what new physics for the complex Berry phase and Berry curvature, let us consider an electric and magnetic fields applied on a condensed matter system. The electric field is perpendicular to the magnetic field, based on the fluctuation-dissipative theorem with the current-current correlation function, the current density of a filled magnetic band is given by\cite{Bohm}
\begin{equation}\label{HJ1}
\mathbf{J}=\frac{e^2}{\hbar}\int_{\textrm{BZ}} \frac{d^{3}k}{(2\pi)^3}\mathbf{\Omega}_{n}\times \mathbf{E},
\end{equation}
where $\mathbf{\Omega}_{n}$ is the Berry curvature.
For a 2D Hall system with a constant electric field along the $x$ direction, $E_{x}$, the Hall current density is defined by $J_{y}=\sigma_{H}E_{x}$, where the Hall conductance can be expressed by $\sigma_{H}=\frac{e^2}{h}\gamma_{-}$, where $\gamma_{-}$ is the Berry phase. Note that for non-Hermitian systems,
$\Omega_{-}^{z,n}$ is complex. Thus, the Hall conductance for Hermitian systems is generalized to the Hall admittance,
\begin{equation}\label{HC2}
Y=\sigma_{H}+iB,
\end{equation}
where $\sigma_{H}=\frac{e^2}{h}\gamma_{-}^{r}$ is the quantum Hall conductance, where $\gamma_{-}^{r}$ is the real part of the Berry phase, which corresponds to the Hermitian Hamiltonian.
$B=\frac{e^2}{h}\gamma_{-}^{i}$ is the quantum Hall susceptance, where $\gamma_{-}^{i}$ are the imaginary parts of the Berry phase in (\ref{CN1}), which comes from the non-Hermitian effect of the Hamiltonian.
A positive susceptance $B>0$ implies that the system has electric capacitive properties and a negative susceptance $B<0$ implies electric inductive properties. Thus, because of the complex energy bands of the non-Hermitian systems, the quantum Hall conductance is generalized to quantum Hall admittance.

Physically,  the energy bands for the non-Hermitian systems in general are complex, which involves non equilibrium processes, such as energy and probability exchange between system and environment. These dissipative effects lead to an intrinsic susceptance. Interestingly, the non-Hermitian parameters can tune the susceptance to be positive or negative, which corresponds to the electric capacitive or inductive properties.

We can rewrite the susceptance in terms of the capacitance and inductance respectively, namely, $B_{C}=\omega C$ and $B_{L}=-\frac{1}{\omega L}$, where
\begin{eqnarray}\label{CL}
C_{\gamma} &=& \frac{e^{2}}{\hbar\omega}\gamma^{I}_{-}   \\
L_{\gamma} &=& \frac{\hbar}{e^{2} \omega}\frac{1}{\gamma^{I}_{-}}
\end{eqnarray}
are the quantum capacitance and inductance respectively, where $\omega$ is the effective frequency of the system. Consequently, the capacitance and inductance are expressed in terms of the Berry phase. In other words, we give the quantum and geometric effects of the capacitance and inductance for non-Hermitian systems.

\section{Non-Hermitian Dirac model}
\subsection{Energy band structure}
As a typical example, we consider a two-dimensional (2D) non-Hermitian Dirac model,
in which the Zeeman-like magnetic field $\mathbf{h}(\mathbf{k})=\mathbf{k}+i\mathbf{\kappa}$ is also regarded as a generalized complex crystal momentum in the first Brillouin zone(BZ), of which $\mathbf{k}=(k_{x},k_{y},m)$ and $\mathbf{\kappa}=(\kappa_x,\kappa_y,\delta)$.
The $\mathbf{\kappa}$ breaks the Hermiticity and  the time reversal symmetry of the Hamiltonian.
The energy bands are obtained as \cite{Fu}
\begin{equation}\label{EE1}
E_{\pm}= \pm\sqrt{k^{2}-\kappa^{2}+2i\mathbf{k}\cdot\mathbf{\kappa}},
\end{equation}
where $k=\sqrt{k_{x}^2+k_{y}^{2}+m^2}$, $\kappa=\sqrt{\kappa_{x}^2+\kappa_{y}^{2}+\delta^2}$.
In general, the roots $E_{\pm}$ can be rewritten as the sum of the real and imaginary parts, $E_{\pm}=\pm\frac{1}{\sqrt{2}}(E^{r}+\mu iE^{i})$, where
$\mu=+1$ for $\mathbf{k}\cdot\mathbf{\kappa}>0$ and $\mu=-1$  for $\mathbf{k}\cdot\mathbf{\kappa}<0$, respectively, yielding

\begin{eqnarray}\label{KK0}
E^{r} &=& \sqrt{k^{2}-\kappa^{2}+\sqrt{(k^{2}-\kappa^{2})^{2}+4(\mathbf{k}\cdot\mathbf{\kappa})^{2}}} \\
E^{i} &=& \sqrt{-(k^{2}-\kappa^{2})+\sqrt{(k^{2}-\kappa^{2})^{2}+4(\mathbf{k}\cdot\mathbf{\kappa})^{2}}}.
\end{eqnarray}

In the polar representation, the the energy bands can be expressed as $E_{\pm}=\pm E e^{i\mu\phi}$, where

\begin{eqnarray}
E &=&\left[(k^{2}-\kappa^{2})^{2}+4(\mathbf{k}\cdot\mathbf{\kappa})^{2}\right]^{1/4}\\
\phi &=& \arctan \left(\frac{\sqrt{-1+\sqrt{1+\eta^{2}}}}{\sqrt{1+\sqrt{1+\eta^{2}}}} \right),
\end{eqnarray}
where $\eta :=\frac{2\mathbf{k}\cdot\mathbf{\kappa}}{k^2-\kappa^2}$ and the angle $\phi$ describes the phase of the energy bands.
The energy bands of non-Hermitian systems are complex and the energy band structures play an important role in quantum phase of these systems.

In general, the energy gap protects quantum phase as the energy bands deform with the parameters varying, which could associate topological phases.
For a non-Hermitian system the complex energy bands leads to two kinds of energy gaps associated with preservation of symmetry.\cite{Kohei}
One is called the point gap defined by a reference point in the complex energy plane, which as an obstacle obstructs the complex energy bands deformation continuously crossing the reference point.\cite{Kohei} The point gap closing with varying the parameters associates with the topological phase transition occurs between the trivial and non-trivial topological phases.
The other is called the line gap defined by the reference line in the complex energy plane, which as an obstacle obstructs the complex energy bands deformation
continuously crossing the reference line. \cite{Kohei} More precisely,
the point gap can be represented by $\det H(\mathbf{k})\neq 0$ in all BZ such that all eigenenergies are nonzero
and the line gap by $\det H(\mathbf{k})\neq 0$ in all BZ and either $E^{r}_{\pm}\neq 0$ or $E^{i}_{\pm}\neq 0$. \cite{Kohei}

When the point gap occurs its energy band structure can be continuously deformed by a unitary flattening transformation and keep their gap structures invariant.
For the line gap, the energy band structures can be continuously deformed by a Hermitian or anti-Hermitian flattening transformation and keep their gap structures invariant. These invariants imply the existence of the topological phase based on the unitary and Hermitian flattening theorems.\cite{Kohei}

Supposing that the BZ is within $-1\leq \left(k_{x},k_{y}\right)\leq 1$ (here we ignore $\pi$ for convenience),
$\det H(\mathbf{k})\neq 0$ leads to $k  \neq  \kappa$ or $\mathbf{k \cdot\kappa} \neq 0$ such that the point or line gap occurs.

Let us investigate the energy band structures of a few typical cases. Fig.1 shows two gap cases, in which (a) and (b) both of the real and imaginary gaps occur. In (c) there is a real band gap, but the imaginary band gap closes and inverses (has a $\pi$ phase flip), which corresponds to $\mathbf{k \cdot\kappa} > 0$ and $\mathbf{k \cdot\kappa} < 0$.
The structure of the energy band gap can preserves under the unitary or Hermitian flattening transformation to associate with the topological phase. \cite{Kohei}

\begin{figure}[!t]\label{Fig1}
\begin{center}
\includegraphics[scale=0.4]{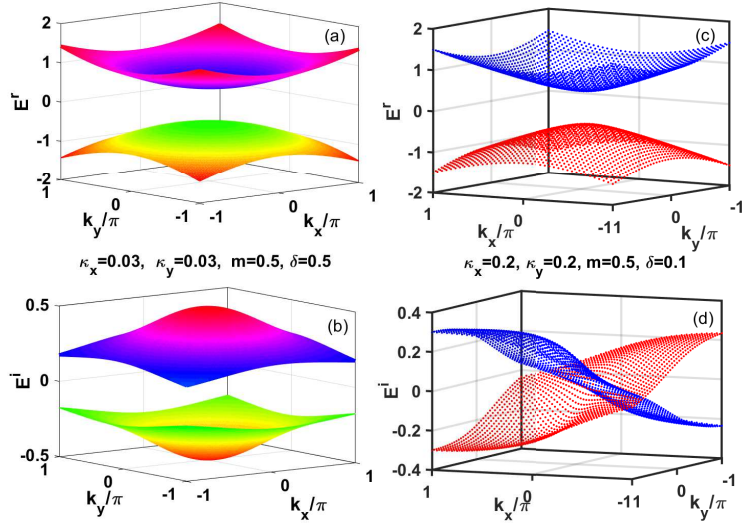}
\caption{
Online color: The complex energy band of Bloch electrons in the BZ, in which (a) and (b) for small $\kappa_x=\kappa_y=0.03$, the real and imaginary parts of the bands shows a gap; (c) when $\kappa_x$ and $\kappa_y$ are tuned to $0.2$ the real part of the band still remains gap, (d) but the imaginary parts of the bands inverse with a line as expected from Eq.(\ref{kkk1}).}
\end{center}
\end{figure}

When the band gaps close as the parameters vary the topological phases will reduce to the trivial phases. Thus, the gap closings in the BZ
as exceptional degenerate states are the critical points or curves of the topological phase transition and they satisfy two equations,

\begin{eqnarray}\label{kkk1}
k_{x}^{2}+k_{y}^{2}-\kappa_{x}^{2}-\kappa_{y}^{2}&=& \delta^2-m^{2} \\
k_{x}\kappa_{x}+k_{y}\kappa_{y} &=& -m\delta.
\end{eqnarray}
When the parameters satisfy the Eq.(\ref{kkk1}) both of the energy band gaps $E^{\textrm{r}}_{\pm}$ and $E^{\textrm{i}}_{\pm}$ close and the topological phases disappear.\cite{Kohei}
We tune the parameters and show a few typical cases in Fig. 2.  We can see that the Dirac cone of the real energy band occur and  the imaginary bands inverses along the line $k_{y}=k_{x}$ in the BZ in Fig. 2 (a) and (b) for $\delta=0$. Inversely, in Fig.2 (c) and (d) for $m=0$, the real band gap closes and inverses along the line $k_{y}=k_{x}$ in the BZ.
The imaginary band touch near $\Sigma$ points but open near $\Gamma$  point in the BZ as expected from Eq.(\ref{kkk1}).
These gap closings as corresponding exceptional degenerate states play a reference-state role for the point or line gap to associate with the topological phase. \cite{Kohei}
Consequently, the necessary condition of the existence of band gap is $m\delta\neq 0$.

We plot a few particular cases in Fig. 3 to illustrate the simultaneous solutions of the Eq.(\ref{kkk1}) where the energy band gaps closing.

    (a) When $\kappa_{y}=0$,   $k_{x}=-m\delta/\kappa_x$, the gap closes at $(k_{x},k_{y})=(-m\delta/\kappa_x,0)$ which corresponds to two lines $(k_{x}=-m\delta/\kappa_x>0$ or $<0$) parallel to the $k_{y}$ axis.

    (b) When $\kappa_{x}=\kappa_{y}$ it leads to the gap closes at $(k_{x},k_{y})=(0,\pm m\delta/\kappa_y)$ and
        and $(k_{x},k_{y})=(\pm m\delta/\kappa_x,0)$. These two parallel lines have angles $3\pi/4$ to the $k_{x}$ axis.

    (c) When $m=0$ or $\delta=0$, the real band gap closes at $(k_{x},k_{y})=\sqrt{2}(\kappa^{2}-m^2,\kappa^{2}-m^2)$.

\begin{figure}[!t]\label{Fig2}
\begin{center}
\includegraphics[scale=0.4]{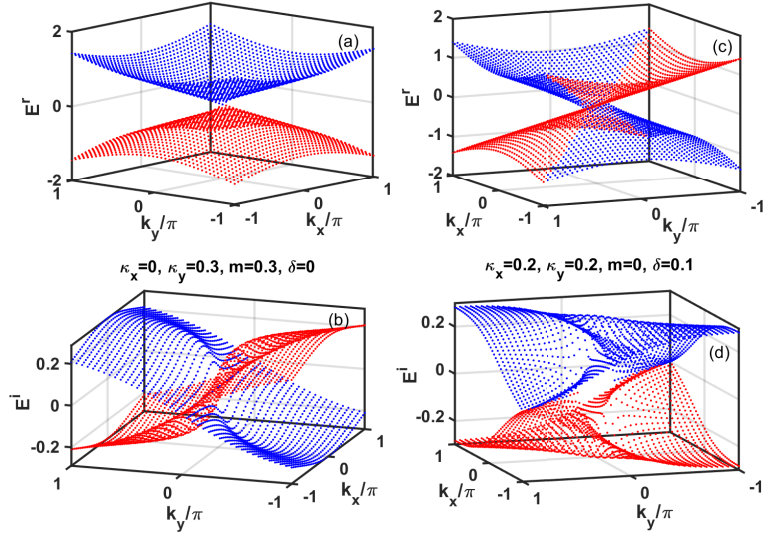}
\caption{
Online color: The complex energy band in BZ, in which (a) the real part of the bands shows a Dirac cone; (b) the imaginary parts of the bands inverse; (c) the real bands inverse along the $k_{y}=k_{x}$ namely $\frac{\pi}{4}$ in BZ, which is called a Weyl node corresponding to $m=0$; (d) the imaginary part of the bands closes at the $\Sigma$ point at the BZ.}
\end{center}
\end{figure}

\begin{figure}[!t]\label{Fig3}
\begin{center}
\includegraphics[scale=0.3]{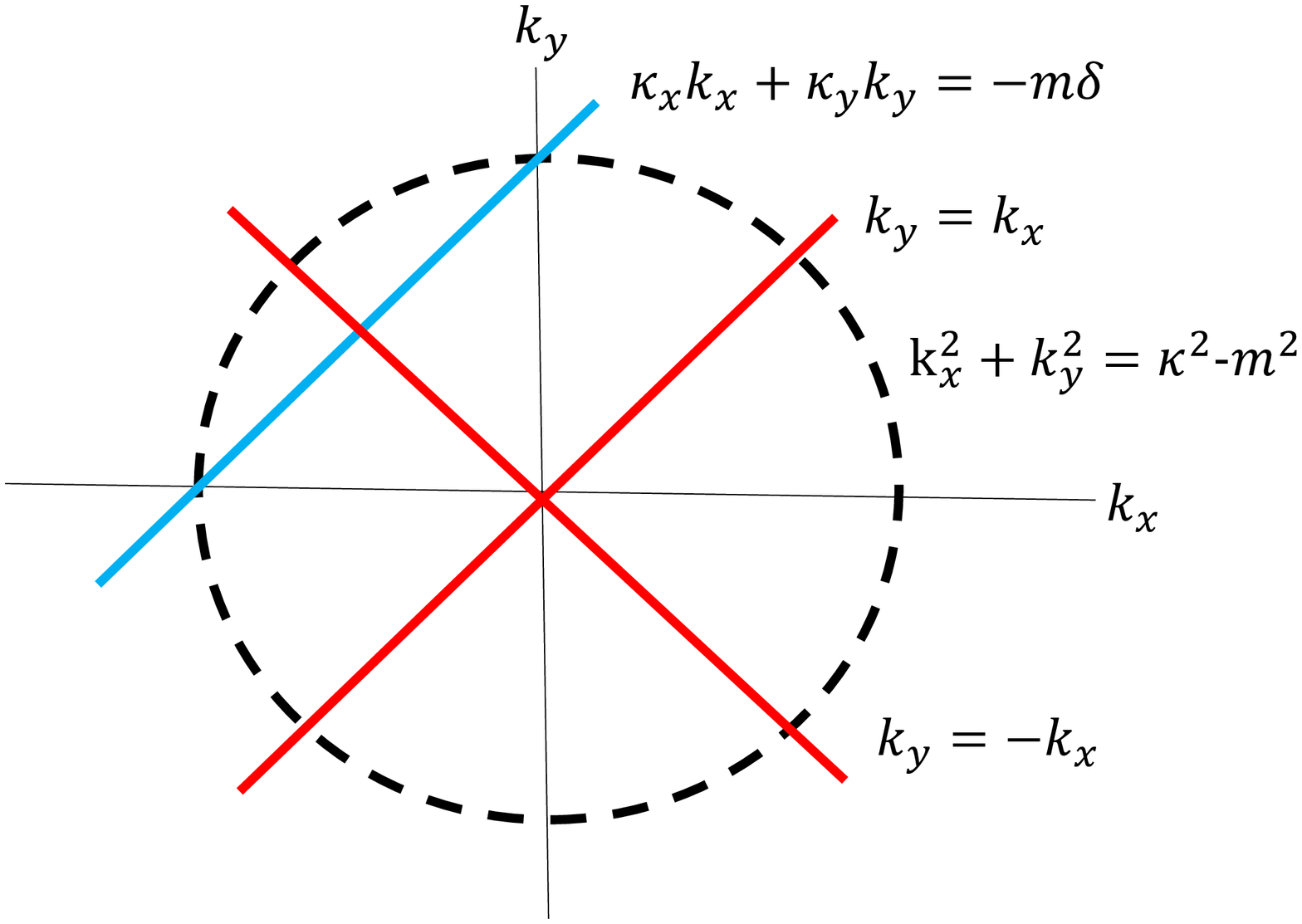}
\caption{
Online color: The typical cases of energy gap closure in the BZ where the cross points of the circle and lines are the critical points that enable the energy gap to close when the parameters satisfy Eq. (\ref{kkk1}).}
\end{center}
\end{figure}

Interestingly, both of the real and imaginary energy bands could change a $\pi$ phase (band inversion) at the line $\mathbf{k\cdot\kappa}=0$, which is a critical line changing the interplay between the energy gaining and losing and a reference line to associate with the topological phase for non-Hermitian systems.

\subsection{Complex Berry connection, Berry curvature and Berry phase}
Suppose that the simultaneous eigen equations of the 2D non-Hermitian Dirac model are written as
\begin{equation}
\mathcal{H}|\psi_{\pm}^R\rangle = E_{\pm}|\psi_{\pm}^R\rangle, \quad
\mathcal{H}^{\dagger}|\phi_{\pm}^L\rangle = E_{\pm}^{*}|\phi_{\pm}^L\rangle.
\end{equation}
The normalized wave functions are given by

\begin{eqnarray}\label{WF1}
|\psi_{\pm}^{R}\rangle &=& \frac{1}{2E_{\pm}(E_{\pm}\pm h_{z})}\left(
\begin{array}{c}
h_{x}- ih_{y} \\
E_{\pm}-h_{z}
\end{array}
\right) \\
|\phi_{\pm}^{L} \rangle &=& \frac{1}{2E_{\pm}^{*}(E_{\pm}^{*}\pm h_{z})}\left(
\begin{array}{c}
h_{x}^{*}- ih_{y}^{*} \\
E_{\pm}^{*}-h_{z}^{*}
\end{array}
\right),
\end{eqnarray}
which forms a complete biorthonormal basis,
$\left\{\psi_{\pm}^R,\phi_{\pm}^L\right\}$ and $\left\langle \phi_{\alpha}^L|\psi_{\beta}^R\right\rangle=\delta_{\alpha\beta}$, where $\alpha,\beta=\pm$.
The complex Berry connection is defined by
$A_{\pm} = \langle \phi_{\pm}^{L}|\nabla_{\mathbf{k}} \mathcal{H}|\psi_{\pm}^R\rangle \cdot d\mathbf{k}$ and we can give
\begin{equation}\label{CBC2}
A_{\pm}=\frac{-h_{y}dk_{x}}{2E_{\pm}(E_{\pm}-h_{z})}+\frac{h_{x} dk_{y}}{2E_{\pm}(E_{\pm}-h_{z})}.
\end{equation}
Note that $\widehat{\mathbf{h}}\cdot
\left(\frac{\partial \widehat{\mathbf{h}}}{\partial k_{x}}\times
\frac{\partial \widehat{\mathbf{h}}}{\partial k_{y}}\right)=\frac{h_{z}}{E^3}$,
we can obtain the complex Berry curvatures for the occupied band,
\begin{equation}\label{BC5}
\Omega_{-}^{z} = \frac{h_{z}}{2E^3},    \quad   \Omega_{-}^{z\dagger}=\frac{h_{z}^{*}}{2E^{*3}},
\end{equation}
where $E=\sqrt{k^{2}-\kappa^{2}+2i\mathbf{k}\cdot\mathbf{\kappa}}$. By some algebra, the complex Berry curvatures can be simplified to

\begin{eqnarray}\label{CN3}
\Omega_{-}^{z,r} &=& \frac{mE^{r3}-\delta E^{i3}-3mE^{r}E^{i2}+3\delta E^{r2}E^{i}}{E^3} \\
\Omega_{-}^{z,i} &=& \frac{mE^{i3}+\delta E^{r3}-3mE^{r2}E^{i}-3\delta E^{r}E^{i2}}{E^3}.
\end{eqnarray}

It can be seen that both real and imaginary Berry curvatures acquire a singularity at the exceptional point $E=0$.
Let us investigate the behavior of the complex Berry curvature at the exceptional points.
When the parameters satisfy the circle equation in Eq.(\ref{kkk1}), $k_{x}^{2}+k_{y}^{2}=\kappa^2-m^2$, the complex Berry curvature is reduced to
\begin{equation}\label{CBC2}
\Omega_{-}^{z,r} =\Omega_{-}^{z,i} = \frac{-m+\delta}{4(\mathbf{k\cdot\kappa})^{3/2}}.
\end{equation}
Consequently, when $m=\delta$, the Berry curvature vanishes $\Omega_{-}^{z,r} =\Omega_{-}^{z,i} =0$. However, when $m\neq\delta$ and $\mathbf{k\cdot\kappa}=0$
both real and imaginary parts of the complex Berry curvature are divergent, namely, the exceptional points of the solution of Eq.(\ref{kkk1}) are singularity for the Berry curvature, which implies the topological phase emergence.

When $\mathbf{k\cdot\kappa}=0$, namely $\phi=0$,  the complex Berry curvature is reduced to
\begin{equation}\label{CBC3}
\Omega_{-}^{z} = \frac{m+i\delta}{2(k^{2}-\kappa^{2})^{3/2}} \\
\end{equation}
Similarly, at the exceptional points, the cross points of the circle and lines in (\ref{kkk1}), both real and imaginary parts of the complex Berry curvature are divergent.
When $\kappa=\delta=0$, the complex Berry curvature reduces to be real $\Omega_{-}^{z}=\frac{m}{2k^{3}}$, and consistent with the previous result.\cite{Andrei,Ryu}
The exceptional degeneracy plays a defect role in the energy band to form a topological phase labeled by winding number, which implies the topological phase depends on the non-Hermitian parameters.

The energy band structures near the exceptional points play an important role. The energy of Bloch electrons is no longer conserved for non-Hermitian systems, which yields some interesting physical properties.

To explore the low-energy topological properties of the model, we study the Berry phase of the model in the continuum limit of the long-wavelength regime, namely we extend the integral region to infinity. Thus we obtain the Berry phase,
\begin{eqnarray}\label{CN9}
\gamma_{-} &=& -\frac{i}{2\pi}\frac{m+i\delta}{\sqrt{(m+i\delta)^2}} \left[ \ln \left(\frac{\kappa_x+m+i\delta}{\kappa_x-m-i\delta}\right) \right. \nonumber\\
   & + & \left. \ln \left(\frac{-\kappa_y-(m+i\delta)(\kappa_x-(m+i\delta))+i\kappa_y\sqrt{-\kappa_{x}^{2}-\kappa_{y}^{2}+(m+i\delta)^{2}}}
{-\kappa_y+(m+i\delta)(\kappa_x+(m+i\delta))+i\kappa_y\sqrt{-\kappa_{x}^{2}-\kappa_{y}^{2}+(m+i\delta)^{2}}}\right)\right],
\end{eqnarray}
where the factor on the left of the logarithm will turn out $\pm 1$ that depend on $m$ or $\delta$. In general, the  Berry phase is complex.
We analyze some special cases.

When $\kappa_x=0$ or $\kappa_y=0$, the  Berry phase becomes
\begin{equation}\label{CN10}
\gamma_{-} = \frac{m+i\delta}{2\sqrt{(m+i\delta)^2}}=\left\{
\begin{array}{c}
  \frac{1}{2} \frac{m}{|m|}, \quad  \textrm{for} \quad \delta=0 \\
  \frac{1}{2} \frac{\delta}{|\delta|}, \quad  \textrm{for} \quad m=0.
\end{array}
\right.
\end{equation}
The Berry phase reduces to $\gamma_{-}=\pm\frac{1}{2}$ once
$\kappa_x=0$ or $\kappa_y=0$, which is consistent with the previous results. \cite{Andrei, Ryu, Ghatak}
Interestingly, either $\kappa_x=0$ or $\kappa_y=0$ the Berry phase is real and positive or negative one half, which depends only on the positive or negative $m$ and $\delta$
whereas is independent of the values of $\kappa_x=0$, $\kappa_y=0$, $m$ and $\delta$. This implies a topological property of quantum Hall effect even though there exists some non-Hermitian effects.

When $m=0$, the Berry phase can be expressed as
\begin{equation}\label{CN11}
\gamma_{-} =\frac{1}{\pi}\frac{\delta}{|\delta|} \left[\arctan \left(\frac{\delta}{\kappa_x}\right)
+\arctan \left(\frac{\delta \kappa_x}{\kappa_{y}^{2}+\delta^2+\kappa_{y}\sqrt{\kappa_{x}^{2}+\kappa_{y}^{2}+\delta^2}}\right)\right].
\end{equation}
The Berry phase becomes real and depends on the non-Hermitian parameters. When $\kappa_x\rightarrow \pm 0$ the Berry phase shows a step with $\pm \frac{1}{2}$, which associates with the different topological phases. When the real energy gap closing ($m=0$) the quantum Hall effect depends on the non-Hermitian parameters.

When $\delta=0$, the Berry phase can be obtained by
\begin{equation}\label{CN12}
\gamma_{-} =
\left\{
\begin{array}{l}
-\frac{i}{2} \frac{m}{|m|}
\left[\ln\left(\frac{\kappa_x+m}{\kappa_x-m} \right)+
 \ln\left(\frac{\kappa_{y}^{2}+m(\kappa_x-m)+\kappa_{y}\sqrt{\kappa_{x}^{2}+\kappa_{y}^{2}-m^2}}
{\kappa_{y}^{2}-m(\kappa_x-m)+\kappa_{y}\sqrt{\kappa_{x}^{2}+\kappa_{y}^{2}-m^2}}\right)\right], \quad  \textrm{for} \quad \kappa_{x}^{2}+\kappa_{y}^{2}\geq m^2 \\
-\frac{i}{2} \frac{m}{|m|}\left[
\ln\left(\frac{\kappa_x+m}{\kappa_x-m} \right)
+\ln\left(\frac{\left[m^{2}(m^2-\kappa_{x}^{2}-\kappa_{y}^{2})-\kappa_{x}^{2}\kappa_{y}^{2}\right]^{2}
+4m^{2}\kappa_{x}^{2}\kappa_{y}^{2}\left(m^{2}-\kappa_{x}^{2}-\kappa_{y}^{2}\right)}
{\left[\kappa_{y}^{2}-m(\kappa_x+m)\right]^{2}+\kappa_{y}^{2}(m^2-\kappa_{x}^{2}-\kappa_{y}^{2})}\right)\right]\\
+\frac{1}{2} \frac{m}{|m|} \arctan \left(\frac{2m\kappa_{x}\kappa_{y}\sqrt{m^2-\kappa_{x}^{2}-\kappa_{y}^{2}}}{m^{2}(m^2-\kappa_{x}^{2}-\kappa_{y}^{2})
-\kappa_{x}^{2}\kappa_{y}^{2}}\right),
\qquad \qquad \qquad \qquad \quad \textrm{for}\quad \kappa_{x}^{2}+\kappa_{y}^{2}<m^2.
\end{array}
\right.
\end{equation}
The Berry becomes complex and depends on the non-Hermitian parameter. The quantum Hall conductance depends on the real part of the Berry phase, whereas the quantum Hall susceptance depends on the imaginary part of the Berry phase. The positive imaginary part of the Berry implies the system exhibiting the electric capacitance and the negative imaginary part of the Berry implies the system showing the electric inductance.

In general, the quantum Hall conductance depends on the Berry phase for Hermitian systems. For non-Hermtian systems, the Berry phase becomes complex such that the quantum Hall conductance is extended to the complex domain, namely admittance (\ref{HC2}).

\begin{figure}[!t]\label{Fig4}
\begin{center}
\includegraphics[scale=0.35]{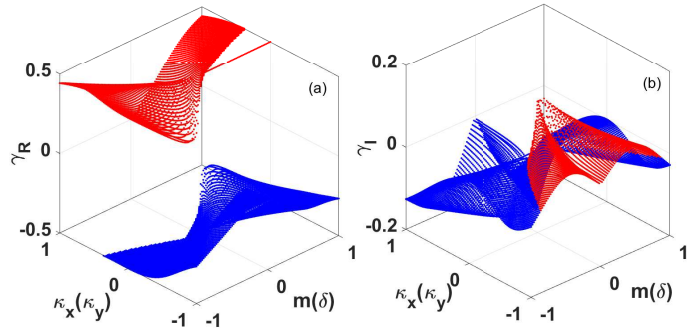}
\caption{
Online color: The complex Berry phase versus the non-Hermitian parameters. (a) The real part of Berry phase. (b) The imaginary part of Berry phase, where $\kappa_x=\kappa_y$ and $m=\delta$. The blue one is negative and the red one is positive.}
\end{center}
\end{figure}

Fig. 4 shows the complex Berry phase in the parameter space based on (\ref{CN9}), in which we set $\kappa_x=\kappa_y$ and $m=\delta$ for convenience as some typical cases.
The real part of the Berry phase in (a) are positive (red part) and negative (blue part) for the values of $\kappa_x(\kappa_y)$ to be positive and negative, whereas the imaginary part of the complex Berry phase are inversely negative and positive for the values of $\kappa_x(\kappa_y)$ to be positive and negative. These results tell us the relationship between the non-Hermitian parameters and quantum Hall admittance, in which the positive and negative imaginary Berry phases correspond to the electric capacitance and inductance respectively. This implies that the electric properties of non-Hermitian systems are transferable between the electric capacitance and inductance by tuning non-Hermitian parameters. This gives us a clue for potential applications in nanoelectronics.

It should be remarked that even though the bulk states characterized by the complex Berry phase cannot capture directly the edge modes due to the non-Hermiticity of systems breaking the bulk-boundary correspondence for Hermitian systems, the bulk-boundary correspondence for non-Hermitian systems can be recovered by a similarity transformation.\cite{Yao} In other words, the bulk states characterized by the complex Berry phase can still trace the edge modes for non-Hermitian systems.

\section{Conclusions and outlooks}
In summary, the geometric and topological properties of non-Hermitian systems exhibit novel quantum phenomena.\cite{Gong,Kohei,Fu}.
The exceptional degeneracy of the energy bands indicates rich new physics of potential relevance for both fundamental concepts and applications.\cite{Ramy,Zhang,Wehling,Tang}
In particular, quantum states of non-Hermitian systems can be classified to different topological classes based on symmetry and topology.\cite{Gong,Kohei,Fu}
A natural question is what new physics behind the topological phase in non-Hermitian systems.

We have proposed a pair of the complex Berry curvature coming from the non-Hermitian Hamiltonian and its Hermitian adjoint
to reveal new phenomena in non-Hermitian systems. We have given the complex Berry curvature and  Berry phase for the 2D non-Hermitian Dirac model and found
that the imaginary part of the Berry phase can induce the quantum Hall susceptance such that the quantum Hall conductance is generalized to the quantum Hall admittance.
The quantum Hall conductance was found to deviate from a quantized value for a specific non-Hermitian Dirac Model.\cite{Yu} Our results are further extended the fractional quantized Hall conductance to the complex domain, namely quantum Hall admittance.
When the non-Hermitian parameter vanishes the quantum Hall admittance will be reduced to the quantum Hall conductance which is consistent with the previous results.\cite{Andrei, Ryu, Ghatak} These results imply that the non-Hermiticity of systems can induce an intrinsic quantum capacitance or quantum inductance respectively associated with positive or negative values of the electric susceptance, depending on the Berry phase and non-Hermitian parameters. Interestingly, this intrinsic quantum Hall susceptance induced by the non-Hermitian effects is beyond the common electric susceptance realized by device design (e.g. in capacitors or inductors), which will inspire us some hints to explore some fundamental physics of non-Hermitian quantum mechanics and its potential applications.

Actually, the quantum transport problem in non-Hermitian system has been studied as a nonequilibrium stochastic dynamics.\cite{Ren,Vladimir}
The nonequilibrium stochastic currents in non-Hermitian systems shows topological nontrivial phases and their topological phase transitions can be classified by braid group and winding number.\cite{Ren,Vladimir} The quantized and fractal quantized currents in stochastic systems depend on the vector potential,\cite{Vladimir} which implies some connections to the quantum Hall admittance we reveal in non-Hermitian Dirac model. The quantum transport in non-Hermitian systems manifests some geometric and topological properties, which promises potential applications.

As a typical example, we have analyzed the complex energy band structures of the 2D non-Hermitian Dirac model and given the energy band gaps and their corresponding Dirac cone and Weyl node as the exceptional degeneracy of the energy bands in the BZ, which are associated with the topological phase. These results demonstrate the predictions of the point and line gaps associated with the topological phase based on the unitary flattening and Hermitian flattening theorems on the point- and line-gap features for non-Hermitian systems.\cite{Kohei} In the continuum limit, we have given the complex Berry phase in the parameter space. The quantized and non-analytic behaviors of the
Berry phase in the parameter space are associated with the topological phases or quantum phase transition.
These findings reveal novel phenomena in non-Hermitian systems and inspire further exploration of their applications in designing quantum devices.

\acknowledgments
The authors thank Dr. Matthew J. Lake for his improving the English presentation of the manuscript.

\begin{appendix}

\subsection{Derivation of Berry curvature of non-Hermitian systems}
Let us consider a generic non-Hermitian quantum system, its Hamiltonian $\mathcal{H}^{\dagger}\neq \mathcal{H}$ acts on an N-dimensional separable Hilbert space. The Schr\"{o}dinger equation is
\begin{eqnarray}\label{SChE}
i\hbar\frac{\partial}{\partial t}|\Psi\rangle & = & \mathcal{H}|\Psi\rangle \\
i\hbar\frac{\partial}{\partial t}|\Phi\rangle & = & \mathcal{H}^{\dagger}|\Phi\rangle.
\end{eqnarray}
For the 2D non-Hermitian Dirac model in the crystal momentum space,\cite{Fu}
\begin{equation}\label{AH1}
\mathcal{H}(\mathbf{k})=\mathbf{h}(\mathbf{k})\cdot \mathbf{\sigma},
\end{equation}
where is the generalized Zeeman magnetic field and $\mathbf{\sigma}$ is the spin-1/2 Pauli operator.
In the adiabatic approximation, the simultaneous eigen equations are written as
\begin{equation}
\mathcal{H}|\psi_{\pm}^R\rangle = E_{\pm}|\psi_{\pm}^R\rangle, \quad
\mathcal{H}^{\dagger}|\phi_{\pm}^L\rangle = E_{\pm}^{*}|\phi_{\pm}^L\rangle.
\end{equation}
The sequences of eigenvectors $\left\{\psi_{\pm}^R,\phi_{\pm}^L\right\}$ form a complete biorthonormal basis, $\left\langle \phi_{\alpha}^L|\psi_{\beta}^R\right\rangle=\delta_{\alpha\beta}$, where $\alpha,\beta=\pm$.
The completeness relation for $\mathcal{H}$ is expressed as
\begin{equation}\label{CR1}
I=P_{+}+P_{-},
\end{equation}
where $P_{\pm}=|\psi_{\pm}^R\left\rangle\right\langle \phi_{\pm}^L|$
are the projection operators. The wave function can be expanded as
\begin{equation}\label{WF1}
|\Psi\rangle=c_{+}|\psi_{+}^R\rangle+c_{-}|\psi_{-}^R\rangle.
\end{equation}
Substituting the wave function to the Schr\"{o}dinger equation (\ref{SChE}) and taking the inner product by acting with the bra $\langle\phi_{\pm}|$, we have
\begin{equation}\label{SchE2}
\frac{d}{dt}\left(
\begin{array}{c}
c_{+} \\
c_{-} \\
\end{array}
\right)=\left(
\begin{array}{cc}
iA_{++}+\frac{1}{i\hbar}H_{++} &  iA_{+-}+\frac{1}{i\hbar}H_{+-}\\
iA_{-+}+\frac{1}{i\hbar}H_{-+} &  iA_{--}+\frac{1}{i\hbar}H_{--}\\
\end{array}
\right)\left(
\begin{array}{c}
c_{+} \\
c_{-} \\
\end{array}
\right),
\end{equation}
where
\begin{equation}\label{AH}
A_{\pm\pm}=i\left\langle \phi_{\pm}^L\left|\frac{d}{dt}\right|\psi_{\pm}^R\right\rangle, \quad
H_{\pm\pm}=\langle \phi_{\pm}^L|\mathcal{H}|\psi_{\pm}^R\rangle.
\end{equation}
In the adiabatic approximation, the off-diagonal elements tend to zero, $\left\langle \phi_{\mp}^L\left|\frac{d}{dt}\right|\psi_{\pm}^R\right\rangle\rightarrow 0$ and $\langle \phi_{\mp}^L|H|\psi_{\pm}^R\rangle\rightarrow 0$.
The adiabatic evolution of states as time varies in the range $0<t\leq\tau$ can be mapped to the boundary of the BZ,
$ -1\leq \left(k_{x},k_{y}\right)\leq 1$.
The wave function of the Bloch electron states can be expressed as
\begin{equation}\label{WF1}
|\psi_{\pm}^{R}(\tau)\rangle=c_{\pm}(0)e^{i\gamma^{B}_{\pm}}e^{i\gamma^{D}_{\pm}}|\psi_{\pm}^{R}(0)\rangle.
\end{equation}
where $\gamma^{B}_{\pm}$ is the complex Berry phase and $\gamma^{D}_{\pm}$ is the complex dynamical phase,
\begin{equation}\label{BP1}
\gamma^{B}_{\pm}=\oint_{\partial(\textrm{BZ})} A_{\pm}, \quad \textrm{and} \quad \gamma^{D}_{\pm}=\int_{0}^{\tau} E_{\pm}d\tau,
\end{equation}
Here, $A_{\pm}=i\left\langle \phi_{\pm}^{L}|d|\psi_{\pm}^{R}\right\rangle$ is the generalized Berry connection for the non Hermitian system. The imaginary parts of $\gamma^{B}_{\pm}$ and $\gamma^{D}_{\pm}$ describe the phenomena of energy dissipation and the non-conservation of probability. The complex Berry curvature is defined as $\Omega_{\pm}=dA_{\pm}$, yielding
\begin{equation}\label{BC11}
\Omega_{\pm}= i\left\langle d\phi_{\pm}^{L}|\wedge|d\psi_{\pm}^{R}\right\rangle= \Omega_{\pm}^{z}dk_{x}\wedge dk_{y},
\end{equation}
where
\begin{equation}\label{BC12}
\Omega_{\pm}^{z}=i\left[\langle\partial_{k_x}\phi_{\pm}^{L} |\partial_{k_y}\psi_{\pm}^{R}\rangle
- \langle\partial_{k_y}\phi_{\pm}^{L} |\partial_{k_x}\psi_{\pm}^{R}\rangle\right].
\end{equation}
Note that $\partial_{k_\alpha}P_{\pm}=|\partial_{k_\alpha}\psi_{\pm}^{R}\rangle\langle\phi_{\pm}^{L}|+|\psi_{\pm}^{R}\rangle\langle\partial_{k_\alpha}\phi_{\pm}^{L}| $, so that
\begin{eqnarray}\label{PP}
\left[\partial_{k_x}P_{\pm}, \partial_{k_y}P_{\pm}\right] &=&|\partial_{k_x}\psi_{\pm}^{R}\rangle\langle\phi_{\pm}^{L}|\partial_{k_y}\psi_{\pm}^{R}\rangle\langle\phi_{\pm}^{L}|
-|\partial_{k_y}\psi_{\pm}^{R}\rangle\langle\phi_{\pm}^{L}|\partial_{k_x}\psi_{\pm}^{R}\rangle\langle\phi_{\pm}^{L}|  \nonumber \\
&+& |\partial_{k_x}\psi_{\pm}^{R}\rangle\langle\partial_{k_y}\phi_{\pm}^{L}|
-|\partial_{k_y}\psi_{\pm}^{R}\rangle\langle\partial_{k_x}\phi_{\pm}^{L}| \nonumber \\
&+& |\psi_{\pm}^{R}\rangle\langle\partial_{k_x}\phi_{\pm}^{L}|\partial_{k_y}\psi_{\pm}^{R}\rangle\langle\phi_{\pm}^{L}|
-|\psi_{\pm}^{R}\rangle\langle\partial_{k_y}\phi_{\pm}^{L}|\partial_{k_x}\psi_{\pm}^{R}\rangle\langle\phi_{\pm}^{L}|  \nonumber \\
&+& |\psi_{\pm}^{R}\rangle\langle\partial_{k_x}\phi_{\pm}^{L}|\psi_{\pm}^{R}\rangle\langle\partial_{k_y}\phi_{\pm}^{L}|
-|\psi_{\pm}^{R}\rangle\langle\partial_{k_x}\phi_{\pm}^{L}|\psi_{\pm}^{R}\rangle\langle\partial_{k_y}\phi_{\pm}^{L}|
\end{eqnarray}
and $\partial_{k_\alpha}\langle\phi_{\pm}^{L}|\psi_{\pm}^{R}\rangle=0$, where $\alpha=x,y$.
The trace is defined as $\textrm{Tr}(\bullet):=\sum_{\pm}\langle\phi_{\pm}^{L}|\bullet|\psi_{\pm}^{R}\rangle$ for the biorthonormal basis.
We then have
\begin{equation}\label{PP2}
\textrm{Tr}\left(P_{\pm}\left[\partial_{k_x}P_{\pm}, \partial_{k_y}P_{\pm}\right]\right)
=\langle\partial_{k_x}\phi_{\pm}^{L}|\partial_{k_y}\psi_{\pm}^{R}\rangle
- \langle\partial_{k_y}\phi_{\pm}^{L}|\partial_{k_x}\psi_{\pm}^{R}\rangle.
\end{equation}
Thus, the Berry curvature can be rewritten as\cite{James}
\begin{equation}\label{BC13}
\Omega_{\pm}^{z}=i\textrm{Tr}\left(P_{\pm}\left[\partial_{k_x}P_{\pm}, \partial_{k_y}P_{\pm}\right]\right).
\end{equation}
In the spectral representation, the Hamiltonian is expressed as
\begin{equation}\label{HH}
\mathcal{H}=E_{+}|\psi_{+}^{R}\left\rangle\right\langle \phi_{+}^{L}|+E_{-}|\psi_{-}^{R}\left\rangle\right\langle \phi_{-}^{L}|.
\end{equation}
Using the completeness relation (\ref{CR1}) and
the spectral representation of the Hamiltonian (\ref{HH}), the projection operators can be rewritten as
\begin{equation}\label{PO3}
P_{\pm}=\frac{1}{2}\left(I\pm \widehat{\mathbf{h}}\cdot \mathbf{\sigma}\right).
\end{equation}
where $\widehat{\mathbf{h}}:=\frac{\mathbf{h}}{h}$ is the dimensionless crystal momentum.
Note that $\partial_{k_\alpha}P_{\pm}=\pm\frac{1}{2}\partial_{k_\alpha}\widehat{\mathbf{h}}\cdot \mathbf{\sigma}$ and using the mathematical identity,
$(\mathbf{a}\cdot\mathbf{\sigma})(\mathbf{b}\cdot\mathbf{\sigma})=\mathbf{a}\cdot\mathbf{b}+i\mathbf{\sigma}\cdot(\mathbf{a}\times\mathbf{b})$,
we obtain the Berry curvature
\begin{equation}\label{BC14}
\Omega_{-}^{z}=\widehat{\mathbf{h}}\cdot
\left(\frac{\partial \widehat{\mathbf{h}}}{\partial k_{x}}\times
\frac{\partial \widehat{\mathbf{h}}}{\partial k_{y}}\right)
\end{equation}
and the  Berry phase for the non-Hermitian Dirac model is expressed as
\begin{equation}\label{CN3}
\gamma_{-}=\frac{1}{4\pi}\int_{\textrm{BZ}} \widehat{\mathbf{h}}\cdot
\left(\frac{\partial \widehat{\mathbf{h}}}{\partial k_{x}}\times
\frac{\partial \widehat{\mathbf{h}}}{\partial k_{y}}\right)dk_{x}dk_{y}.
\end{equation}
Similarly the Berry curvature and  Berry phase can be obtained for the Hermitian adjoint of the Hamiltonian.

\end{appendix}


\end{document}